\documentclass{mn2e}
\usepackage{aas_macros,graphicx,times,multirow}

\usepackage{graphicx}
\usepackage{subfig}
\usepackage{multirow}
\usepackage{array}

\newcommand{\xmm}{{\em XMM-Newton}}
\newcommand{\chan}{{\em Chandra}}
\newcommand{\cxo}{{CXO\, J085201.4$-$461753}}

\title[The counterpart and nebula of the Vela Jr. CCO]{The nature of the infrared counterpart and of the optical nebula associated with  the Central Compact Object in Vela Jr.\thanks{Based on observations collected at the European Organisation for Astronomical Research in the Southern Hemisphere under ESO programmes 077.D-0764(A), 095.D-0960(A), 098.D-0346(A), 177.D-3023(B)}}
\author[R. P. Mignani, et al. ]
{\parbox{\textwidth}{R. P. Mignani$^{1,2}$\thanks{E-mail: roberto.mignani@inaf.it}, 
A. De Luca$^{1,3}$,
S. Zharikov$^{4}$,
W. Hummel$^{5}$,
W. Becker$^{6,7}$,
A. Pellizzoni$^{8}$
} 
\\ \\
$^{1}$ INAF - Istituto di Astrofisica Spaziale e Fisica Cosmica Milano, via E. Bassini 15, 20133, Milano, Italy\\
$^{2}$ Janusz Gil Institute of Astronomy, University of Zielona G\'ora, ul Szafrana 2, 65-265, Zielona G\'ora, Poland \\
$^{3}$INFN, Sezione di Pavia, via A. Bassi 6, I-27100 Pavia, Italy\\
$^{4}$Observatorio Astronomico Nacional, Instituto de Astronomia, Universidad Nacional Autonoma de Mexico, Ensenada, BC, Mexico\\
$^{5}$ European Southern Observatory, Karl Schwarzschild-Str. 2, D-85748, Garching, Germany \\
$^{6}$ Max-Planck Institut f\"ur extraterrestrische Physik, Giessenbachstrasse 1, 85741 Garching, Germany\\
$^{7}$ Max-Planck Institut f\"ur Radioastronomie, Auf dem H\"ugel 69, 53121 Bonn, Germany \\
$^{8}$ INAF-Osservatorio Astronomico di Cagliari, Via della Scienza 5, 09047 Selargius, Italy 
 }

\begin{document}

\date{Accepted 1988 December 15. Received 1988 December 14; in original form 1988 October 11}

\pagerange{\pageref{firstpage}--\pageref{lastpage}} \pubyear{2002}

\maketitle

\label{firstpage}

\begin{abstract}
The X-ray source \cxo\ in the few kyr-old Vela Jr. supernova remnant (G266.2-1.2) belongs to the peculiar class of isolated neutron stars dubbed "Central Compact Objects" (CCOs). At variance with the other CCOs,  which are only detected in the X-rays, \cxo\  has been possibly detected also at other wavelengths.  In particular, a candidate near-infrared counterpart (H = 21.6 $\pm$0.1) was detected by the Very Large Telescope (VLT) in addition to a 6\arcsec-wide nebula detected in H$\alpha$, interpreted as a velocity-driven bow-shock.
Here, we present new near-infrared and optical VLT observations of the candidate counterpart to \cxo\ and its nebula to confirm the proposed associations. Moreover, we used archival \chan\ observations to measure for the first time the \cxo\ proper motion.  The comparison between the two sets of near-infrared VLT images, taken 10.56 year apart, do not show evidence of proper motion for the candidate counterpart to \cxo, expected from its 4\arcmin\ offset from the SNR geometrical centre, with a $3\sigma$ upper  limit of  $\sim 10$ mas yr$^{-1}$. This is much smaller than the expected  proper motion of $\sim$50--100 mas yr$^{-1}$, which, in turn, is below the $3\sigma$ upper limit of $\sim$ 300 mas yr$^{-1}$ that we obtained with \chan. At the same time, VLT spectroscopy resolved the nebula optical emission, which is dominated by the NII doublet at 6548 and 6584 \AA\ and not by H$\alpha$.
To summarise, we conclude that the proposed near-infrared counterpart is  likely unassociated with \cxo\ and that the nebula is not a velocity-driven bow-shock.
\end{abstract}


\section{Introduction}

The idea that rotation-powered pulsars (RPPs) are not the only manifestation of isolated neutron stars (INSs) has now been consolidated by the last 40 years of observations. At least five different INS classes have been recognised, see Harding (2013) for a recent review. Among these classes, one of the most enigmatic is that of the so-called Central Compact Objects (CCOs; Pavlov et al.\ 2000), which owe their name to their association with the central regions of young (a few kyr old) supernova remnants (SNRs). About a dozen of CCOs have been now identified (see De Luca 2017 for a recent review), with the {\em ante litteram} CCO, the X-ray source 1E\, 161348$-$5055 discovered by the {\em Einstein} X-ray Observatory at the centre of the RCW\, 103 SNR (Tuohy \& Garmire 1980), now recognised to be a slowly-rotating magnetar (e.g., Rea et al.\ 2016).  

Despite their presumably young age, CCOs feature characteristics which are very much at variance with those of young RPPs.
First of all, 
they are all undetected in radio. Then, they are only detected through thermal X-ray emission from the neutron star surface with no evidence  of rotation-powered non-thermal emission from the magnetosphere at any wavelength.
Furthermore, at variance with young RPPs they are not embedded in bright pulsar wind nebulae (PWNe). Finally, in the three cases where X-ray pulsations have been detected (Zavlin et al.\ 2000; Gotthelf \& Halpern 2005; Gotthelf et al.\, 2009), the spin periods $P_{\rm s}$ (a few hundreds of ms) and their first derivative $\dot{P_{\rm s}}$ (of the order of $10^{-17}$ s s$^{-1}$; Halpern \& Gotthelf 2010; De Luca et al.\ 2012; Gotthelf et al.\, 2013) point at characteristic ages of a few hundreds of Myr, which are a factor of $\sim 10^5$ larger than those of the associated SNRs, to a very low spin-down energy ($\dot{E} \sim 10^{31}$--$10^{32}$ erg s$^{-1}$), and  to dipole magnetic fields of the order of a few $10^{10}$ G, a factor of $\sim$ 100 lower than those of young RPPs and a factor of 1000--10000 lower than those of the magnetars. The low magnetic field values, as opposed to their young age,  lead to nickname CCOs as "anti-magnetars" (Gotthelf et al.\ 2013).
Why CCOs are so different from other young INSs is still a matter of debate. 
It has been proposed that the properties of CCOs be due to prompt accretion of $10^{-4}$--$10^{-3}$ $M_{\odot}$ fallback material soon after the SN explosion, which would bury the magnetic field of the newborn neutron star. Deep optical/near-infrared observations did not find any evidence for any debris disk still surrounding CCOs a few thousand yr after their birth (Mignani et al.\ 2008, 2009a; De Luca et al.\ 2008, 2011).

One of  the most intriguing CCOs is CXO\, J085201.4$-$461753 (Pavlov et al.\ 2001), discovered by the {\em Advanced Satellite for Cosmology and Astrophysics} (Slane et al.\ 2001)
at the centre of the G266.2$-$1.2 SNR, a.k.a. Vela Jr. (Aschenbach 1998).    The reason is that, at variance with all the other CCOs, it features possible evidence of emission in the optical and near-infrared.  Indeed, using the Wide Field Imager (WFI) at
the 2.2 m MPG telescope at La Silla (European Southern Observatory, ESO) and photographic plates from the  3.9 m UK  Schmidt  Telescope  (UKST) at  the  Anglo-Australian  Observatory (AAO), Pellizzoni et al.\ (2002) discovered relatively bright H$\alpha$ emission from a nebula ($\sim 6\arcsec$ diameter)  at the \cxo\ position. This nebula was tentatively 
interpreted  
as evidence of hydrogen ionisation 
in a velocity-driven bow-shock formed by the CCO wind as it moves 
in the interstellar medium (ISM) 
at a small angle to the line of sight. Such bow-shocks have been indeed observed around some fast-moving neutron stars  (e.g., Brownsberger \& Romani 2014). The nebula  was also detected in the R band (Mignani et al.\ 2007) with the ESO Very Large Telescope (VLT).
 Surprisingly, it was not detected in  {\em Hubble Space Telescope} ({\em HST}) H$\alpha$ images taken with the Wide Field and Planetary Camera 2 (WFPC2), down to a flux $\sim$10 times fainter than expected (Mignani et al.\ 2009b). This could be explained by 
a red-shift of the H$\alpha$ line, still seen through the ground-based filters which are broader and redder than the {\em HST}/WFPC2 one, with the CCO moving  away at a radial velocity of 450--2700 km s$^{-1}$ (Mignani et al.\ 2009b).
At the same time, Mignani et al.\ (2007) also found a possible point-like counterpart (H = 21.6 $\pm$0.1) to the CCO in VLT near-infrared observations, with a position compatible with the \chan\ coordinates. At the SNR distance (1 kpc; Slane et al.\  2001), the object's H--K colour and brightness would be compatible with emission from a disc, a very low-mass companion (M-type or later), or the neutron star magnetosphere. 

Both associations have not been confirmed yet. Therefore, we carried out follow-up observations with the VLT to confirm the CCO  identification in the near-infrared and determine the nature of the nebula. At the same time, we used archival \chan\ observations to measure the \cxo\ proper motion for the first time.  In this manuscript we describe the observations in Sectn.\ 2, with the results presented and discussed in Sectn.\  3 and 4, respectively.

\section{Observations and data reduction}

\subsection{Very Large Telescope}

\subsubsection{NACO Observations}

The star density in the \cxo\ field gives a $\sim2\%$ chance coincidence probability for the proposed counterpart (Mignani et al.\ 2007), making the association uncertain. Following a well-tested approach (e.g., Mignani et al.\ 2000, 2002), this can be confirmed by measuring the proper motion  of the candidate counterpart 
and comparing it in magnitude and direction with that expected for the neutron star.  Indeed, the  $\approx 4\arcmin$ offset due northwest (position angle $\sim 356^{\circ}$ east of north) between the \chan\ position of \cxo\ (Pavlov et al.\ 2011; Mignani et al.\ 2007) and the  geometrical centre of Vela Jr. inferred from the {\em ROSAT} All sky Survey (RASS) images (Aschenbach 1998) implies, for a SNR age of $\sim$ 3  kyr (Slane et al.\ 2001), a proper motion of $\sim$ 80 mas yr$^{-1}$. 
Such a proper motion can be measured for the \cxo\ candidate counterpart through adaptive optics (AO) high spatial resolution near-infrared astrometry. 

In Mignani et al.\ (2007) we obtained a very accurate position of the proposed near-infrared counterpart (epoch May 23 2006) using  NAos COnica (NACO), the AO imager and spectrometer mounted at the 8.2m VLT/UT1 (Lenzen et al.\ 2003; Rousset et al.\ 2003). We obtained second-epoch observations in December 14, 20, 28 2016, with exactly the same instrument set-up as in the first-epoch observation for a direct comparison. In particular, we used 
the  $S27$ camera  ($28\arcsec\times28\arcsec$ field of view, 0\farcs027/pixel) with the VIS dichroic and wavefront sensor and the FowlerNsamp readout mode. The new observations were also obtained  in the $H$ band, where our target was detected with the highest signal--to--noise.    Owing to the recent problems with NACO second quadrant\footnote{{\tt www.eso.org/sci/facilities/paranal/instruments/naco}} we had to offset the pointing by $\Delta\alpha=+5\arcsec$ and $\Delta\delta=-5\arcsec$  to centre our target in the fourth quadrant.   For the AO correction  we used the  same reference  star  as used in  the  first-epoch observation,  S1331311130291 ($V=15.3$), at 11\farcs3 from our target.  Three $H$-band observation blocks (OBs) of 2280 s each (120 s DIT)
    were executed in service mode, in grey time, clear sky conditions, with seeing 0\farcs5--0\farcs8, and with the target close to the zenith (airmass $\sim$1.1).

Night (twilight flat fields) and day time-calibration frames (darks, lamp flat fields) were taken daily as part of the NACO calibration plan. Like in the case of the first-epoch data, we processed the second-epoch ones using the ESO NACO pipeline not to introduce any systematics. We co-added the science images from the single OBs  using the {\sc eclipse} software (Devillard 2001) to increase the signal--to--noise and the accuracy on the target position. Given the small epoch difference between the observations (14 days at most) the proper motion expected for our target ($\sim$ 80 mas yr$^{-1}$) would only introduce a maximum uncertainty of $\sim 0.1$ pixel on its centroid determination.   We computed the astrometry calibration using a set of 2MASS stars (Skrutskie et al.\ 2006) yielding an overall uncertainty of $\sim 0\farcs1$.  For the flux calibration we used the night zero point computed by the NACO pipeline from the observation of standard stars.

\subsubsection{FORS2 Observations}

 The nature of the nebula observed at the \cxo\ position (Pellizzoni et al.\ 2002) is uncertain. As explained in Mignani et al.\ (2009b), the interpretation of a red-shifted H$\alpha$ emission from a velocity-driven bow-shock  competes with that of [NII] emission from an unrelated knot of gas, peraphs  associated with the planetary nebula (PN) candidate Wray 16-30 (Reynoso et al.\ 2006) $\sim$25\arcsec\ to the southwest or with a complex of bright arc-like structures seen $\sim$ 30\arcsec--40\arcsec\ to the west.  High-resolution optical spectroscopy of the nebula is, then, key to discriminate between the possible interpretations.  Since X-ray pulsations have not been detected yet  (Kargaltsev et al.\  2002; Becker et al.\ 2006) confirming the bow-shock scenario would be important to constrain the CCO spin-down energy $\dot{E}$ (Pellizzoni et al.\ 2002).
 
 We observed the nebula with the FOcal Reducer and low dispersion Spectrograph (FORS2; Appenzeller et al.\ 1998), also installed at the VLT/UT1. We observed  in long-slit spectroscopy (LSS) mode.   We used the high-resolution grism GRIS\_1200R+93 with central wavelength $\lambda_{\rm C}$=6500 \AA\ ($\lambda_{\rm min}$--$\lambda_{\rm max}$=
5750--7310 \AA) and a resolving power $\lambda$/$\Delta$$\lambda$=2140 at the central wavelength.  We used the filter GG435+81 and  the standard-resolution collimator (0\farcs25/pixel). In order to  better cover the central (and brighter) part of the nebula  we used a slit of  1\arcsec\ width.  We chose this set-up  to achieve  an effective spectral resolution of 3 \AA/pixel around the H$\alpha$ rest wavelength (6562 \AA) and detect the presumed red-shift of the H$\alpha$ line. The slit size also helps to minimise the possible contamination from the wings of the bright  Star Z ($R\sim 17$; Pavlov et al.\ 2001) located $\approx 4\arcsec$ away from the nebula (see Fig.\ 1 of Pellizzoni et al.\ 2002). The slit was aligned eastwest with a small offset of -1\farcs3 in declination to include a bright reference star $\sim$1\arcsec\ east, which was used for the blind offset required to position our target at the centre of the slit. Three OBs of 2760 s each, split in two exposures for cosmic ray filtering, were executed in service mode  on May 14 and 15 2015 in grey time and under clear sky conditions. The seeing was mostly sub-arcsec, with an average value of $\sim 0\farcs7$, and the airmass was 1.1--1.2.

We reduced and calibrated the LSS spectra with the dedicated tools in the {\sc IRAF}\footnote{IRAF is distributed by the National Optical 
Astronomy Observatories, which are operated by the Association of Universities for Research in Astronomy, Inc., under cooperative agreement with the National Science Foundation.}  and {\sc MIDAS}\footnote{{\tt https://www.eso.org/sci/software/esomidas/}} software packages. The standard data reduction includes bias subtraction, flat-fielding, cosmic-ray removal, and bad pixel correction.  We then summed the six  two-dimensional spectra from the three OBs and computed the wavelength calibration from the spectra of calibration lamps using the {\tt twodspec} task in IRAF.  We extracted the one-dimensional spectrum at the target position using  a rectangular window of 4 pixel width (1\arcsec) and we subtracted the sky background and the sky lines  sampled in a region 3\arcmin\ to the east  of our target not to be affected by the contribution of extended structures or bright stars in the field of view (Fig. \ref{vst}a). We computed the flux calibration with the task 
{\tt onedspec} in IRAF using spectra of the spectrophotometric standard star LTT3218 observed during  the same nights as our target. We extracted the standard star spectrum using a window of the same size as used for our target.

\subsection{VLT Survey Telescope}

To complement the VLT optical spectroscopy, we used serendipitous multi-band images of the field of \cxo\ obtained with the Omegacam 4$\times$8 CCD array   (Arnaboldi et al.\ 1998; Kuijken et al.\ 2002; Kuijken 2011) at the 2.6m VLT Survey Telescope (VST).  The images were taken on May 11 2015 as part of the VST Photometric H-Alpha Survey of the Southern Galactic Plane and Bulge (VPHAS+; Drew et al.\ 2014).   The narrow-band H$\alpha$ filter NB\_659 (4$\times$120s)  and the Sloan filters u\_sdss (2$\times$150s), g\_sdss (4$\times$40s), r\_sdss  (6$\times$25s), i\_sdss  (3$\times$25s)  were used, with the numbers of exposures and exposure times reported in parenthesis. Owing to the large Omegacam field of view ($1^{\circ} \times 1^{\circ}$; 0\farcs21/pixel) the filters are segmented in four quadrants (A,B,C,D) in the SW to NW direction (clockwise) each covering a 2$\times$4 CCD sub-array. Due to dithering, \cxo\ falls in different CCDs in different exposures, hence in different filter quadrants. For instance, in the H$\alpha$ exposures it falls in CCDs \# 24, 28, 5, 7 corresponding to quadrants B, C, and A (Drew et al.\ 2014). For all filters, the central wavelengths and widths of the four quadrants differ from each other by a few \AA. For the NB\_659 filter we assume an average central wavelength $\lambda_{\rm C}$=6589.3 \AA\  and
FWHM=102.7\AA.
 The night was in clear sky conditions and the seeing during the exposures was $\sim 0\farcs9$--$1\farcs3$ and the airmass 1.2--1.3. 
 
 Single images were reduced by the VPHAS+ pipeline run at the Cambridge Astronomical Survey Unit (CASU), which applies bias correction, flat fielding, astrometry and photometry calibration, and retrieved from the ESO science portal\footnote{{\tt http://archive.eso.org/scienceportal/home}}. Finally, per each filter we co-added the single science images with the {\sc Swarp} software (Bertin et al.\ 2002). To minimise the computing overheads we only co-added the CCD chips corresponding to the coordinates of the nebula. The tiny difference in the quadrant-dependent central wavelengths and widths of the Sloan and narrow band filters is not expected to produce a noticeable effect in the co-added images.

\subsection{Chandra}

The \cxo\ proper motion has not been directly measured yet. Since the identification with its proposed near-infrared counterpart (Mignani et al.\ 2007) is still unconfirmed, obtaining an independent proper motion measurement in the X-rays with \chan\ would be essential in many respects.  On one hand,  this would serve as a prime reference for comparison with the proper  motion of the proposed near-infrared counterpart to \cxo, which would be used to directly confirm its identification. On the other hand,  the \chan\ proper motion value would constrain the CCO transverse velocity and distance, whereas its direction would provide independent and undisputed evidence of the association of \cxo\ with the Vela Jr. SNR.  In addition, extrapolating back in time the \chan\ proper motion would provide an independent determination of the birth place of the CCO and of the age of the Vela Jr. SNR.  

We note that  \chan\  already  measured the proper motion of another CCO, RX\, J0822$-$4300 in the Puppis A SNR (Becker et al.\ 2012; Gotthelf et al.\ 2013). The proper motion value, computed over a time baseline of $\sim$10.5 yrs, is  $\mu = 71\pm12$ mas yr$^{-1}$ or $\mu = 61.0\pm8.8$ mas yr$^{-1}$, comparable to that expected for the CCO in Vela Jr. ($\sim 80$ mas yr$^{-1}$),
which gives us confidence that it can be measured too,
given a long-enough time base line.
\cxo\ has been observed three times by \chan\  in Guaranteed Time with both the ACIS (Advanced CCD Imaging Spectrometer) and HRC (High Resolution Camera) instruments. The first time was on October 26 2000 and was a snapshot observations of  3 ks with the ACIS-I (ObsID 1032),  the second was on September 17 2001 with the ACIS-S for 31.49 ks (ObsID 1034) and the last on November 13 2009 with the HRC-I  for 28.26 ks (ObsID 10702). The second observation was performed in continuous clocking mode and it cannot be used for our purpose. The HRC-I and ACIS-I observations were performed in imaging mode with the target on axis and $\sim 2\farcm6$ off-axis, respectively and are suitable for astrometry. The epoch difference between the two observations provides a time base line of $\sim$ 9 years.

We retrieved the data from the \chan\ science archive\footnote{{\tt http://cxc.harvard.edu/cda/}} and reprocessed the event files using the {\tt chandra\_repro} script (version 4.11) of the {\sc Chandra Interactive Analysis of Observations} ({\sc CIAO}) software package\footnote{{\tt http://cxc.harvard.edu/ciao/}}.

\begin{figure*}
\begin{tabular}{cmd}
\subfloat[]{\includegraphics[width=8.cm, angle=0,clip=]{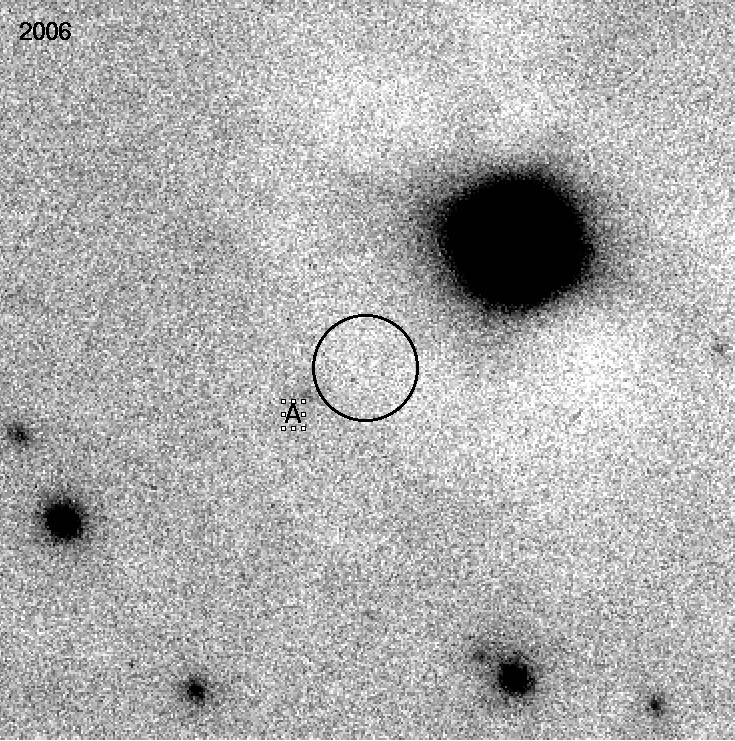}} 
\subfloat[]{\includegraphics[width=8.1cm, angle=0,clip=]{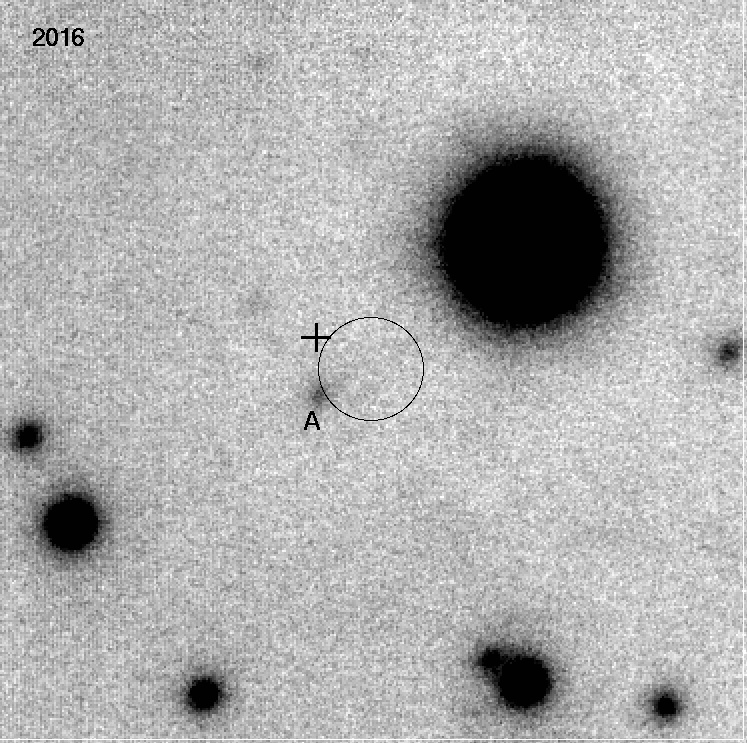}} 
\end{tabular}
\caption{Comparison between $10\arcsec\times10\arcsec$ sections of the two VLT/NACO H-band images of the \cxo\ field taken on May 23 2006 (2280 s) and  December 2016 (6840 s).  North to the top, east to the left. In both panels, the circle (0\farcs7 radius) marks the \chan\ position (epoch 2000.8) as computed in Mignani et al.\ (2007): $\alpha=08^{\rm h} 52^{\rm m}
01\fs37 $, $\delta= -46^\circ 17\arcmin 53\farcs50$. The bright star northwest to the error circle is Star Z of Pavlov et al.\ (2001). The quality of the second-epoch image reflects both the better seeing conditions and the longer integration time. The candidate counterpart to \cxo\ (object A) is labelled as in Mignani et al.\ (2007). The cross in panel b) marks its expected position, 
$\alpha=08^{\rm h}  52^{\rm m} 01\fs44$, $\delta= -46^\circ 17\arcmin 53\farcs08$,  
at the epoch of the second NACO observation (2016.9) computed for a proper motion of 80 mas yr$^{-1}$ and a position angle of $\sim 356^{\circ}$ east of north (Sectn.\ 2.1.1). \label{naco} 
}
\end{figure*}

\section{Results}

\subsection{The CCO }

\subsubsection{The candidate counterpart proper motion}

The 2006 and 2016 NACO H-band images (Sectn.\ 2.1.1.) define the first and second-epoch reference for the proper motion measurement of the \cxo\ candidate counterpart. We used a set of well-suited reference stars in common between the two images to register them on a relative reference frame by applying a coordinate transformation which accounts for the shifts in X and Y directions, rotation angle and plate scale using the tasks {\tt geomap} and {\tt geotran} in {\sc IRAF}. The NACO detector is affected by a time-dependent distortion different for the different cameras but which has been characterised to a high degree of accuracy (see, e.g. Plewa et al.\ 2015). Since the pointing of the second-epoch image has been offset to position the target at the centre of the fourth quadrant, the local distortion map is different than at the cross of the four quadrants, where the target was positioned in our first-epoch image. Moreover, also the AO correction  is different in the first and second-epoch images owing to the different position of the reference star in the detector. This affects the accuracy on the registration of the two images and the determination of the centroids of the reference stars and of our target, owing to variations in the position-dependent PSF. However, since the expected angular displacement of our target between the two epochs is of the order of 30 pixels ($\sim 0\farcs85$), for a proper motion of 80 mas yr$^{-1}$ and a time base line of 10.56 yr, these effects should play a minor role. For instance,  the detector distortion can affect the determination of the projected target position by  $\sim 0.2$ pixels at the cross of the four quadrants and by up to $\sim 0.7$ pixels at the centre of the fourth quadrant.  Therefore,  in the first place we neglect the geometric distortion correction. After the registration process described above, the two frames are aligned with an accuracy of $\sim 0.5$ pixel (rms) in both the X and Y directions, respectively, where the relatively large rms incorporates the effects of the  uncorrected geometric distortion.

The first and second epoch NACO images are shown in Fig. \ref{naco}, aligned in right ascension and declination. As it can be seen, no obvious displacement of the candidate counterpart can be recognised by eye, as it would be, instead, the expected $\sim$30 pixel displacement due northwest. 
To put constraints on the displacement of the candidate counterpart more quantitatively, we computed its detector coordinates  by fitting a PSF to the intensity profile. The coordinates are $x_{\rm 2006}=449.50\pm0.5$, $y_{\rm 2006}=550.50\pm0.5$ and $x_{\rm 2016}=450.06\pm0.2$, $y_{\rm 2016}=550.94\pm0.2$  in the 2006 and 2016 images, respectively. This yields a  displacement $\Delta x=0.56\pm0.53$ and $\Delta y=0.44\pm0.53$ between the two epochs for the candidate counterpart, where the associated errors ($1\sigma$) only account for the accuracy of the PSF fitting. After accounting for the accuracy on the frame registration ($\sim 0.5$ pixel per coordinate) 
the total error on the computed displacement is 
$\sim 0.73$ pixel ($1\sigma$) per coordinate. Therefore, the computed displacement is not significant. We thus set a $3\sigma$ limit on the displacement of the candidate counterpart of 
$\sim 2.7$ and $\sim 2.6$ pixel in the X and Y direction, respectively, corresponding to a total angular displacement  $\la 4$ pixel.
After accounting for the detector plate scale (0\farcs027) and the time span between the two epochs (10.56 yr), this corresponds to a proper motion $\la10$ mas yr$^{-1}$.

\subsubsection{The CCO proper motion}

We  used the two \chan\ ACIS-I (16\farcm9$\times$16\farcm9 field of view; 0\farcs492/pixel) and HRC-I (30\arcmin$\times$30\arcmin; 0\farcs131/pixel) observations, covering a time baseline of $\sim$ 9 years, to measure the \cxo\ proper motion. Unfortunately,  in each data set the CCO is the only X-ray source detected in the detector field of view, so that we could not measure its displacement through relative astrometry. Therefore,  we were left with the only option of measuring the CCO proper motion through absolute astrometry.  We computed the target position in the ACIS-I and HRC-I data sets using the same approach as described in Becker et al.\ (2012). In brief, we used the task {\tt ChaRT}\footnote{{\tt http://cxc.harvard.edu/chart/index.html}} to compute the model PSF at the target positions in the two data sets and then the  \chan\ {\tt Sherpa}\footnote{{\tt http://cxc.harvard.edu/sherpa/}} modelling and fitting package to compute the corresponding best-fit positions in detector coordinates.  As a last step, we converted these values to equatorial sky coordinates ($\alpha,\delta$) applying the \chan\ aspect solution.  

A direct comparison between the \cxo\  coordinates computed in the ACIS-I and HRC-I data sets did not show any obvious offset over the 9 years. Assuming in both cases the nominal 0\farcs8  accuracy (90\% confidence level) on the \chan\ absolute astrometry\footnote{{\tt http://cxc.harvard.edu/cal/ASPECT/celmon/}}, this implies a $3 \sigma$ upper limit of $\sim$ 300 mas yr$^{-1}$ on the \cxo\ proper motion, much larger than the expected one. Nonetheless, this is the first observational constraint on the proper motion of \cxo\  and the fourth time that  a CCO proper motion measurement has been carried out after RX\, J0822$-$4300 in Puppis A (Becker et al.\ 2012; Gotthelf et al.\ 2013), CXOU\, J232327.8+584842 in Cas A (DeLaney \& Satterfield 2013), and 1E\, 1207.4$-$5209 in G296.5+10.0 (Halpern \& Gotthelf 2015). Our measurement might be improved only if deeper \chan\ observations would reveal more X-ray sources to be used as a reference for relative astrometry. Indeed, \xmm\ observations (Becker et al.\ 2006) show a few bright enough X-ray sources (4) within a $\sim 5\arcmin$ radius around \cxo.  
According to our simulations, all of them would be detected in deep \chan\ ACIS-S observations (40 ks), which would give a $3\sigma$ proper motion sensitivity of $\sim 60$ mas yr$^{-1}$ for a time base line of $\sim$10 years.
This means that 
the required observations (at least two) might still be scheduled within the operational lifetime of the \chan\ mission, which can be extended up to 2030.

\subsection{The Nebula}

\begin{figure}
\begin{tabular}{cmd}
\subfloat[]{\includegraphics[width=8.2cm]{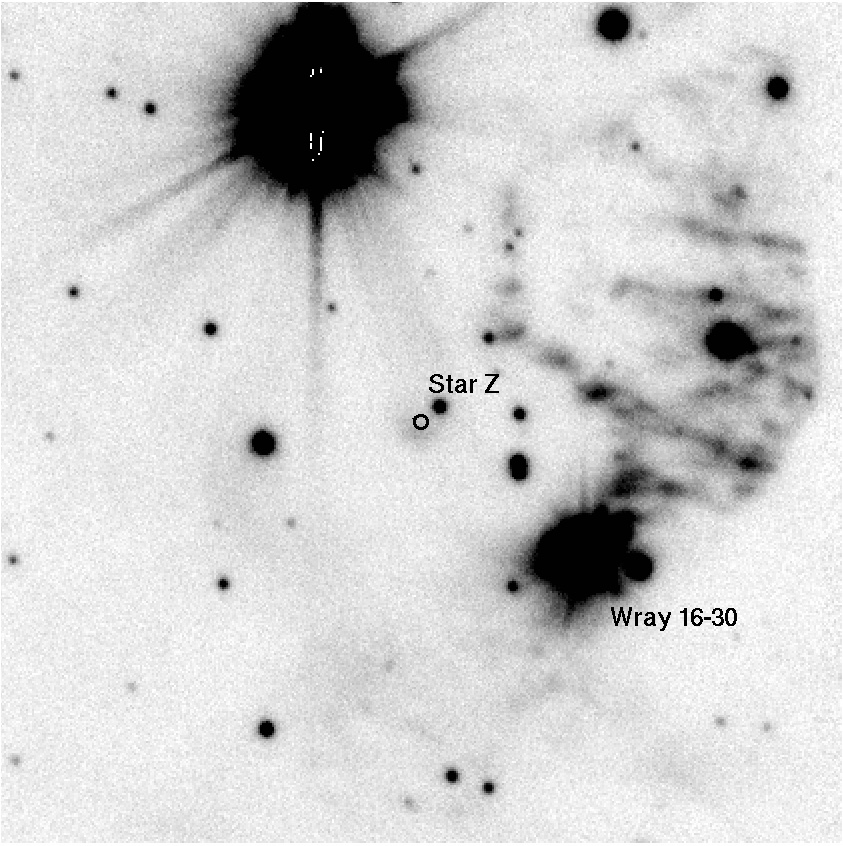}}  \\
\subfloat[]{\includegraphics[width=4.1cm]{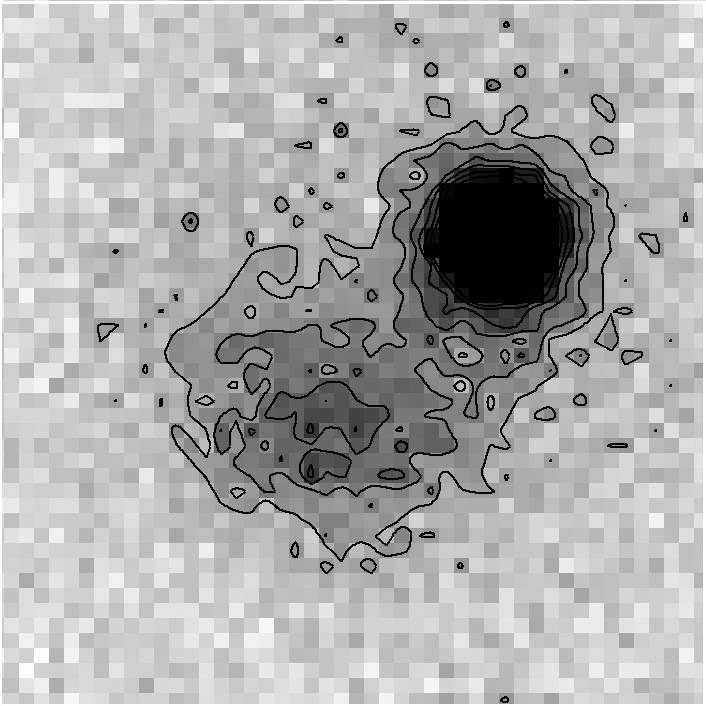}} 
\subfloat[]{\includegraphics[width=4.1cm]{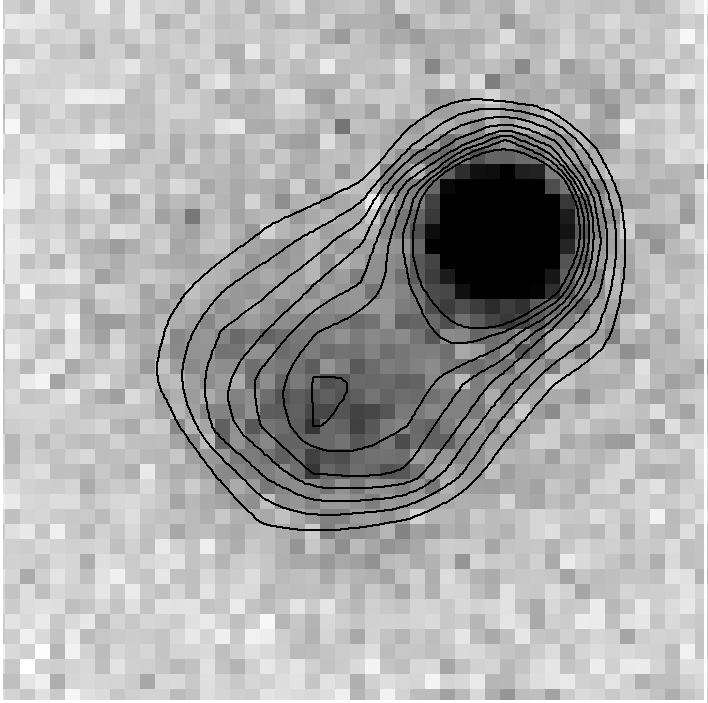}} 
\end{tabular}
\caption{a) Co-added 480s image of a $90\arcsec\times90\arcsec$  region around  \cxo\ obtained with Omegacam on VST in the NB\_659 filter.  North to the top, East to the left. Star Z of Pavlov et al.\ (2001) and the PN candidate  Wray 16-30 (Reynoso et al.\ 2006) are labelled. The  \chan\ position of \cxo\ (Mignani et al.\ 2007) is marked by the circle. The compact nebula (6\arcsec\ diameter) southeast of Star Z and coincident with the \cxo\ position is the bow-shock candidate of Pellizzoni et al.\ (2002).  
 b) Zoom ($10\arcsec\times10\arcsec$) on the nebula with the intensity contours overlaid (linear scale).  c)  Same but with the contours from the UKST H$\alpha$ image of Pellizzoni et al.\ (2002) overlaid.\label{vst} }
\end{figure}

\subsubsection{Multi-band imaging}

Those obtained with the VST are the first multi-band images of the putative \cxo\ nebula obtained so far, which may provide qualitative information on its spectrum. The nebula is not detected in the Sloan Omegacam filters.
It is clearly detected, however, in the 480 s narrow-band H$\alpha$ filter NB\_659 (Fig. \ref{vst}a), 
as expected given the similarity in central wavelength and width with the H$\alpha$ filters of the WFI ($\lambda_{\rm C}$=6588.2 \AA, 
FWHM=74.31\AA) and the UKST ($\lambda_{\rm C}$=6583.5 \AA, 
FWHM=43.6 \AA), 
which are redder and broader  than the {\em HST}/WFPC2 656N filter ($\lambda_{\rm C}$=6564 \AA, 
FWHM=22 \AA) used by Mignani et al.\ (2009b). 
The nebula detection in the narrow-band NB\_659 filter, as opposed to its non-detection in the broad-band ones, seems to be consistent with a very faint (or absent) continuum emission
and a  spectrum that is emission-line dominated\footnote{The nebula was  indeed detected in the R band only thanks to the longer integration (5200 s) and the larger collecting area of the VLT (Mignani et al.\ 2007). }. 
This is the first detection of the nebula in the H$\alpha$ band  since the original observations of Pellizzoni et al.\ (2002), which independently confirms its existence. Furthermore, the Omegacam observations provide the highest spatial resolution H$\alpha$ images  of the nebula and of the complex of bright arc-like structures  observed 30\arcsec--40\arcsec\ West of \cxo\ (Fig.\ref{vst}a) obtained so far. Indeed, the former is not detected in the  {\em HST} images of Mignani et al.\ (2009b) whereas the latter falls only marginally in the WFPC2 mosaiced image. 
These structures are seen through the NB\_659 filter   but not through the broad-band ones, with a marginal detection only in the r\_sdss filter owing to the short integration time\footnote{Indeed, these structures are clearly seen in the R$_{\rm F}$-band images of the DSS-II (Reynoso et al.\ 2006).
} (150 s). This suggests that, like the nebula, they are characterised by an emission-line dominated spectrum.
Other diffuse emission-line regions, but with a lower surface brightness, are also seen south and east of \cxo\ (Fig. \ref{vst}a) as part of a network of large-scale filaments that  extend across the entire Omegacam field--of--view.

In the H$\alpha$ band (Fig.\ref{vst}b) the nebula  is more clearly resolved against Star Z  than in the UKST image, with a clear maximum of emission at the centre of symmetry. This seems to indicate that the two objects are detached and, thus, that the nebula cannot be an unipolar gas outflow from Star Z
(see also, Mignani et al.\ 2009b).
We qualitatively compared the  Omegacam image of the nebula (May 11 2015) with the 
UKST  (November 9 1999) H$\alpha$ image of Pellizzoni et al.\ (2002) to look for possible long-term variations in morphology and/or surface brightness associated with the displacement of the hypothetical bow-shock as \cxo\ moves in the ISM.The comparison is shown in Fig. \ref{vst}c, where we overlaid the intensity contours of the UKST image over the Omegacam one.
As it can be seen, the morphology of the nebula does not show any obvious variation between the two epochs, accounting for the different spatial resolution of the two images. 
No significant variation in the surface brightness can be appreciated either.  Both instances are somehow unexpected for a dynamical structure such as a velocity-driven bow shock (see, e.g. the case of the Guitar Nebula, Chatterjee \& Cordes 2004) but they are more foreseeable  if \cxo\  indeed moves at a small angle to the line of sight.  The morphology of  the arc-like structures West of the nebula looks more fragmentary in the Omegacam image which now distinctly shows substructures, such as filaments,  voids, and knots, some of which seemingly detached from the body of the main structures.   
Also in this case, the comparison with the UKST H$\alpha$ image does not show any obvious variation either in brightness or morphology.

\begin{figure*}
\centering
\begin{tabular}{cmd}
{\includegraphics[width=12cm]{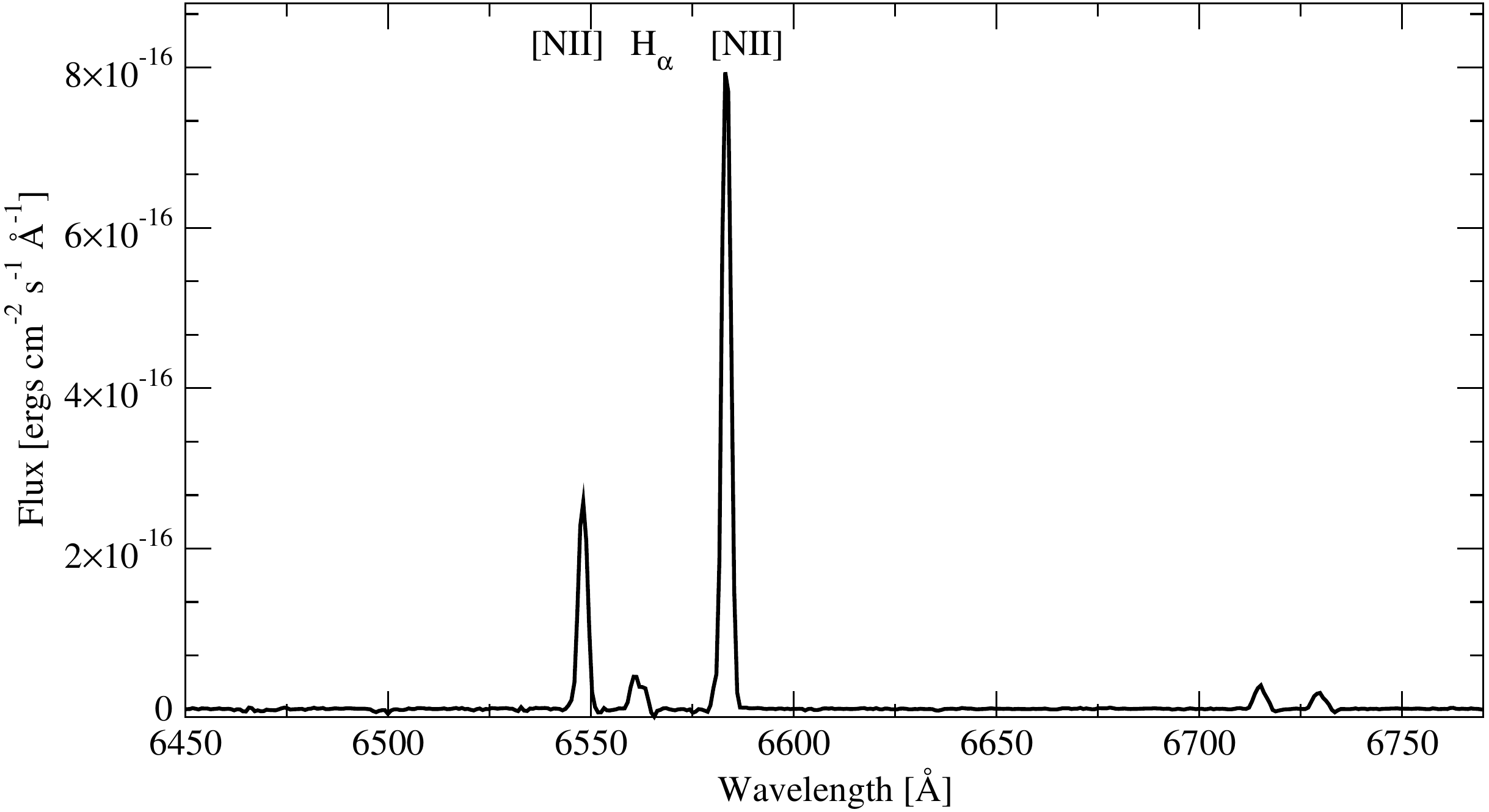}}  \\
\end{tabular}
\caption{VLT/FORS2 spectrum of the nebula associated with \cxo. The two strong emission lines at 6548 \AA\ and 6584 \AA\ are the [NII] doublet. The fainter emission line at 6562.8 \AA\ is H$\alpha$. The two faint emission lines at 6716.47 and 6730.85 \AA\ are part of the [SII]  doublet. \label{fors} 
}
\end{figure*}

\subsubsection{Optical Spectroscopy}

The FORS2 optical spectrum of the nebula is shown in Fig. \ref{fors}. The spectrum is characterised by the absence of continuum emission, as suggested by our multi-band imaging (Sectn.\ 3.2.1), whereas 
around 6500 \AA\ is dominated by two strong and narrow emission lines that we identified as the [NII] doublet at 6548 \AA\ and 6584 \AA, with equivalent width (EW) of  0.46 and 0.72 \AA, respectively.  At longer wavelengths we also identified the [SII]  doublet at 6716.47 and 6730.85 \AA.  The possibility that the brightest  of the two NII lines is associated with a red-shifted H$\alpha$ line is unlikely since the red-shift would have to be exactly the same as the difference between the rest wavelengths of the two lines. Moreover, this would make difficult to explain why the NII only appears as a single line at  6648 \AA. Indeed, we  found evidence of the H$\alpha$ line at its rest wavelength of 6562.8 \AA,  with a relatively broadened profile (EW=81.2\AA). Its intensity is much lower than that of the [NII] doublet, therefore its contribution in the wavelength range of the ground-based H$\alpha$ filters (see Sectn.\ 3.2.1) is marginal. 

The  fact that the spectrum at 6500--6600 \AA\ is dominated by the [NII] doublet explains why the nebula emission was not detected in the   {\em HST}/WFPC2 656N images of Mignani et al.\ (2009b). Indeed, this narrow filter  ($\lambda_{\rm C}$=6564 \AA, FWHM=22 \AA) only covers the wavelength interval between the two NII lines, where the H$\alpha$ emission is too weak to be detected in the co-added WFPC2 images (4$\times$500 s) which have a $3\sigma$ detection limit of $3\times10^{-15}$ erg cm$^{-2}$ s$^{-1}$ (Mignani et al.\ 2009b). In fact, from the integrated H$\alpha$ line intensity measured in the FORS2 spectrum, $I_{6562.8} \sim 0.19 \times10^{-15}$ erg cm$^{-2}$ s$^{-1}$,  and once  the correction for the slit width (1\arcsec) has been applied, we find that the H$\alpha$ flux of the nebula is a factor of three below the WFPC2 656N detection limit.
On the other hand, we find that by combining the integrated intensity of the NII lines in the FORS2 spectrum, $I_{6548} \sim 0.73 \times10^{-15}$ erg cm$^{-2}$ s$^{-1}$ and $I_{6584} \sim 2.2 \times10^{-15}$ erg cm$^{-2}$ s$^{-1}$, again after  the correction for the slit width has been applied, the total flux  is comparable to that measured for the nebula in the UKST H$\alpha$ images ($\lambda_{\rm C}$=6583.5 \AA, FWHM=43.6 \AA), accounting for the difference in the flux calibration\footnote{Pellizzoni et al.\ (2002) computed the flux calibration of the UKST H$\alpha$ image against images of the same field taken in parallel through the Short-Red broad-band filter, which, in turn, were calibrated against the R$_{\rm F}$-band magnitudes from a preliminary release (2.2.0, June 2001) of the GSC-2 catalogue (Lasker et al.\ 2008).}. This confirms that the nebula flux is almost entirely ascribed to emission from the [NII] doublet with only a marginal contribution from the H$\alpha$ line. 
Therefore, we  cannot attribute the nebula origin to pure H$\alpha$ emission from a velocity-driven bow-shock, like it has been for a long time hypothesised (Pellizzoni et al.\ 2002; Mignani et al.\ 2009).  

The absence of strong H$\alpha$ emission  also argues against a possible nebula origin from hydrogen  photo-ionisation in the ISM by ultraviolet radiation from the CCO (Pellizzoni et al.\  2002).
At variance with bow-shocks,  however, 
evidence of such photo-ionisation nebulae
has not been found yet in H$\alpha$ observations of isolated neutron stars  (e.g., Brownsberger \& Romani 2014), suggesting that it is a more rare phenomenon.
Although hydrogen photo-ionisation would be expected in first place,  the presence of strong [NII] lines  in the nebula spectrum leads one to consider whether nitrogen photo-ionisation could be  an alternative possibility. 
Usually, photo-ionisation of metals, such as nitrogen, occurs in gaseous environments where the metal density  is
higher than expected for average ISM conditions.
However, other emission lines that could be tracers of the photo-ionisation process,
such as  [OI] at 6300,6363 \AA, are not detected in the spectrum of the putative \cxo\ nebula, hinting at a local metal density more typical of the average ISM.
Therefore, a possible origin of the nebula [NII] emission by photo-ionisation of nitrogen atoms  is not corroborated by independent evidence that this process indeed occurs in the ISM around \cxo. A comparison between the observed line flux and the predictions of photo-ionisation models would yield to uncertain conclusions since they depend on a number of unknowns, such as the nitrogen ionised fraction and density, as well as on the spectrum of the ionising source and its distance. For \cxo,  the ultraviolet spectrum is unknown, 
having never been observed either by the {\em HST} or the  {\em Galaxy Evolution Explorer}, 
whereas its distance is uncertain by a factor of two (0.5--1 kpc; Allen et al.\ 2015). 
As we discuss later in Sectn.\ 4.3,  other interpretations of the nebula origin seem to be more likely.

\section{Discussion}

\subsection{The CCO identification}

The upper limit on the proper motion of the \cxo\ candidate counterpart (10 mas yr$^{-1}$) is much smaller than the expected value of 80 mas yr$^{-1}$  (Sectn.\ 2.1.1). This value, however, is assumption-dependent since it  has been computed from the $4\arcmin$ offset between the \cxo\ position and the estimated geometrical centre of the SNR (Aschenbach 1998) and for a SNR age of 3 kyrs (Slane et al.\ 2001).  Proper motion estimates inferred in this way can be very uncertain, if not wrong, and must be taken with the due caution. As an example, using \chan\ Halpern \& Gotthelf  (2015) measured  a proper motion of $\mu=(15\pm7$) mas yr$^{-1}$ for the CCO 1E\, 1207.4$-$5209 in the G296.5+10.0 SNR, which is much smaller than the value of $\sim 70$ mas yr$^{-1}$ predicted from the offset of the CCO from the estimated geometrical centre of the SNR and an age of 7 kyr (De Luca et al.\ 2011). 
The SNR age obviously represent the first source of uncertainty in computing the expected proper motion.  Recently, by measuring the Vela Jr. expansion rate with \chan\  Allen et al.\  (2015) determined an age of 2.4--5.1 kyr.  Even for the largest value of the age, however, the 4\arcmin\  offset of \cxo\ from the geometrical centre of the SNR (Aschenbach 1998)  would imply a proper motion of $\sim 50$ mas yr$^{-1}$, still five times larger than the upper limit of $\sim 10$ mas yr$^{-1}$ measured for its candidate counterpart. Thus, a SNR age of $\ga$ 25 kyr would be required to be compatible with the proper motion upper limit, which appears unrealistic.  Indeed, Allen et al.\ (2015) showed that the estimated SNR age might be larger in case of expansion in a non-uniform ISM but only by $\sim$ 50\%. The determination of the SNR geometrical centre, expected to coincide with the neutron star birth place,  represents the second source of uncertainty, which is more difficult to quantify depending on the shape and symmetry of the SNR.  In the case of \cxo,  its actual birth place should be either at $\la$0\farcm4   or at $\la$0\farcm85  from its present position, instead of at 4\arcmin\  (Aschenbach 1998),  to be compatible  with the $\la 10$ mas yr$^{-1}$ proper motion  of its candidate counterpart and the limits on the SNR age (2.4--5.1 kyr). This would imply that either the SN explosion that formed \cxo\ did occur in a place different from what is now the geometrical centre of the SNR, which would be the case for an asymmetric explosion,  or that the centre determination  is wrong by about  3\arcmin.
 Although the almost perfectly circular shape of this young SNR  facilitates the determination of its geometrical centre, a $\sim$3\arcmin\ uncertainty  is still a small fraction of its angular size ($\sim 1^{\circ}$ radius) and is not unrealistic. Therefore, we cannot firmly rule out that the actual geometrical centre of the SNR is closer than claimed to the \cxo\ present position.  
A more precise characterisation of the SNR morphology  though X-ray images at a much higher spatial resolution than that of the RASS ($\sim60\arcsec$)
would help to decrease the uncertainty on the SNR geometrical centre and validate this possibility.  
We conclude that we have no  direct evidence that the object pinpointed by Mignani et al.\ (2007)  is the counterpart to \cxo\  but, at the same time,  no indisputable evidence that it is not, although evidence seem to point in this direction.
The measurement of its proper motion with \chan\  would be a way to firmly rule out the claimed identification. Indeed,  the estimated $3\sigma$  \chan\ proper motion
sensitivity of $\sim 60$ mas yr$^{-1}$  (Sectn.\ 3.1.2)  is obviously incompatible with the $\sim 10$ mas yr$^{-1}$ upper limit for the candidate near-infrared counterpart.
No new candidate counterpart is detected at the \chan\ position in our second-epoch, deeper NACO observations (Fig. \ref{naco}b).  
Thus, \cxo,  like all the other CCOs, still remains unidentified at wavelengths other than the X-rays.

The near-infrared detection limits of  $J \sim  22.6$, $H \sim  22.5$, and $K_s  \sim  21.8$, measured in the first-epoch NACO observation (Mignani et al.\ 2007), are compatible with the emission from an INS, virtually ruling  out an hypothetical late-type stellar  companion to \cxo. The chances of detecting it in the near-infrared or in the optical are difficult to quantify, though. Since no spin period and period derivative have been measured in the X-rays  (Kargaltsev et al.\  2002; Becker et al.\ 2006) we have no constraint on its spin-down energy $\dot{E}$ and we cannot predict its non-thermal near-infrared or optical flux, under the assumption that  the luminosity in both bands scales with $\dot{E}$ the same way as in RPPs (e.g., Mignani et al.\ 2012).  
Our new NACO observations pushed the detection limit down to $H \sim  23.8$ ($3\sigma$), which we assume as the new upper limit on the CCO near-infrared flux.  This value is close to the sensitivity limit of current near-infrared observing facilities, which discourages a new follow-up at these wavelengths.  In the optical, however, we are still far from the limit and 
observations much deeper than the Omegacam ones (Sectn.\ 2.2) might be attempted, although the interstellar reddening to \cxo\ is significant, $E(B-V)\sim 0.67$ (Mignani et al.\ 2007).  
At variance with the near-infrared, where the INS emission is non-thermal, the optical emission can also be ascribed to thermal radiation from the neutron star surface at temperatures of a few $10^{5}$ K (Mignani 2011). 
The only information that we have on the \cxo\ surface temperature has been obtained from the X-ray spectrum. This is modelled by a 
double blackbody  with temperatures of 4.4 and 6$\times10^{6}$ K  (Becker et al.\ 2006) ascribed to radiation emitted from two different areas  (0.36 and 0.08 km radii for a 1 kpc distance) which are too hot and too small to produce detectable optical emission. The temperature of the rest of the neutron star surface, expectedly cooler than the areas responsible for the X-ray emission, remains unconstrained.  Thus,  like in the case of non-thermal emission, we cannot predict the optical flux.  A very deep exploratory observation, e.g.  down to magnitude $\sim$28 in the V band, might be worthwhile to evaluate the chance of success of a multi-band optical follow-up.

\begin{figure*}
\centering
\begin{tabular}{cmd}
\subfloat[]{\includegraphics[bb=15 265 260 520,clip=,height=8cm]{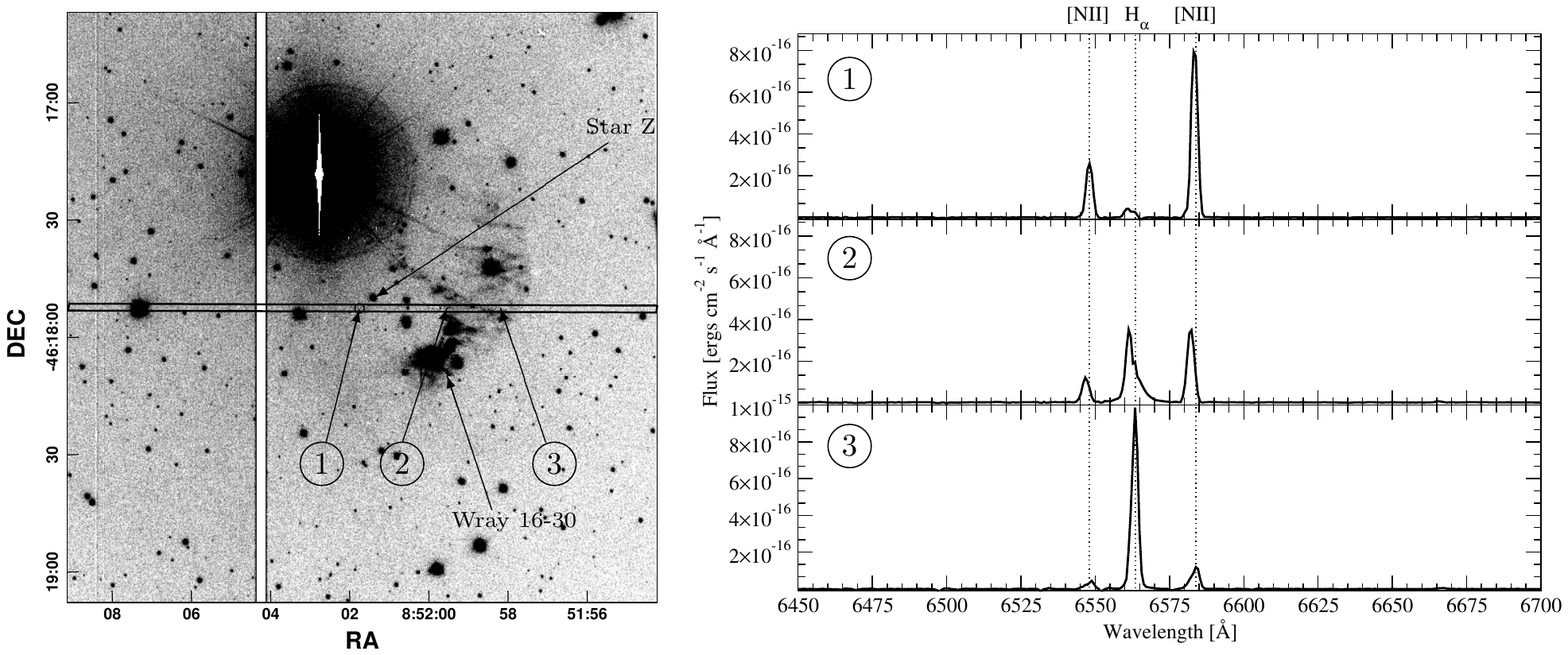}}  
\subfloat[]{\includegraphics[bb=265 265 580 520,clip=,height=8cm]{Spectrum-eps-converted-to.pdf}}  
\end{tabular}
\caption{a) VLT/FORS2  image of the \cxo\ field (R band; 24s) used for target acquisition. The labels (1,2,3) correspond to the positions of the nebula (undetected in the image) and of two regions within the bright arc-like structures detected west of it, respectively. The horizontal lines indicate the slit projection on the plane of the sky. The white vertical stripe corresponds to the gap between the two FORS2 CCDs. Star Z and the PN candidate Wray 16-30 are indicated. The relatively bright star in the slit in the left part of the panel is the reference used  for the blind target acquisition. b) VLT/FORS2 spectra of the nebula and of the two regions defined in panel a).  \label{fors2} 
}
\end{figure*}

\subsection{The CCO velocity}

From the measured \chan\ upper limit on the \cxo\ proper motion (300 mas yr$^{-1}$) we can derive the first direct constraints on its transverse velocity. This can be computed as $V_{\rm T,100}\sim 5\times \mu_{1} \times d_{100}$, where  $V_{\rm T,100}$ is the transverse velocity in units of 100 km s$^{-1}$, $\mu_{1}$ the proper motion of units of 1 arcsec yr$^{-1}$, and $d_{100}$ is the distance in units of 100 pc.
At the estimated SNR distance of $\sim$1 kpc (Slane et al.\ 2001; Allen et al.\ 2015), the upper limit on the proper motion obtained by \chan\ implies a transverse velocity $V_{\rm T} \la 1500$ km s$^{-1}$,  whereas much smaller velocities are inferred from the expected proper motion (Sectn.\ 4.1).  The values of  100 mas yr$^{-1}$ or  $\sim$  50 mas yr$^{-1}$,  for a SNR age of 2.4 and  5.1 kyr (Allen et al.\ 2015) respectively, would imply  $V_{\rm T} \sim 500$ km s$^{-1}$ or $V_{\rm T} \sim 250$ km s$^{-1}$. 
A factor of two smaller transverse velocities are obviously obtained for the lower limit   on the SNR distance of 0.5 kpc (Allen et al.\ 2015). The expected proper motions would, then, imply a transverse velocity  $V_{\rm T} \sim 125$--$500$ km s$^{-1}$ for a distance between 0.5 and 1 kpc, i. e. between the low end and the average of the pulsar velocity distribution (Hobbs et al.\ 2005).
For comparison, the transverse velocity of the CCO RX\, J0822$-$4300 in Puppis A is, for a distance of 2 kpc, $V_{\rm T}=(672\pm115$) km s$^{-1}$  (Becker et al.\ 2012) or  $V_{\rm T}=(619\pm126)$ km s$^{-1}$  for a distance of 2.2$\pm$0.3 kpc (Gotthelf et al.\ 2013).  Transverse velocities smaller by $\sim$60\% are inferred for the 1.3$\pm$0.3 kpc distance measured by Reynoso et al.\ (2017).  These velocities values can be up to a factor of five higher than estimated for \cxo\ from the expected proper motion. 
This seems to suggest that the value of the kick velocity imparted to the neutron star at birth is not a distinctive character for the CCOs and cannot track differences in the dynamic of the supernova explosion with respect to other types of INSs. Only a direct proper motion measurement of \cxo\ with \chan, as well as of other CCOs, would yield an assumption-independent transverse velocity determination which is needed to verify this speculation.  
Apart from  \cxo\ and RX\, J0822$-$4300, proper motion measurements with \chan\ have been carried out only for other two CCOs, CXOU\, J232327.8+584842  in Cas A (DeLaney \& Satterfield 2013) and 1E\, 1207.4$-$5209 in G296.5+10.0 (Halpern \& Gotthelf 2015), yielding $V_{\rm T} \la 790$ km s$^{-1}$ (at 3.4 kpc) and $V_{\rm T} \la 180$ km s$^{-1}$ (at 2 kpc), respectively.  Therefore, no firm conclusion on the CCO velocity distribution can be drawn from the present constraints and just one measurement. Only for  1E\, 1207.4$-$5209 there seems to be evidence of a relatively low transverse velocity. However, the inferred value is not as peculiar as it might seem since also RPPs can have quite low velocities (see, Verbunt et al.\ 2017) like, e.g. the Vela pulsar (PSR\, B0833$-$45) for which $V_{\rm T} \sim 65$ km s$^{-1}$ (Caraveo et al.\ 2001).

\subsection{The Origin of the Nebula}

The absence of strong H$\alpha$ emission in the spectrum of the nebula, which is dominated by the [NII] doublet (6548,6584 \AA),
pushes towards an interpretation different from 
a  velocity-driven bow shock, or  a photo-ionisation nebula, associated with the CCO. 
In this case,  the positional coincidence with  \cxo\ would be spurious, leaving the origin of the nebula an open question.
The region around the CCO is rich  in diffuse emission-line regions, as it can be appreciated from the Omegacam narrow-band 
image (Fig. \ref{vst}a), with the presence of the bright candidate PN Wray 16-30 (Reynoso et al.\ 2006), bright arc-like structures north and north-west of it, and large-scale filaments which cross the entire  field--of--view and intersect close to the \cxo\ position. Therefore, 
the simplest possibility is that 
the nebula
is actually associated with [NII] emission from 
one of these  regions. 
Some possibilities were already discussed in Mignani et al.\ (2009b). 

An identification with an emission knot from the candidate PN Wray 16-30 at $\sim 25\arcsec$, seems to be a likely explanation according to our spectroscopy of the nebula. 
Indeed,  the [NII] doublet at 6548 \AA\ and 6584 \AA\ is always present in PN spectra and the ratio between the intensity of the two lines $I_{6584}/I_{6548}$ is always equal to three, regardless of the physical conditions inside the PN (Gurzadyan 1970), exactly as observed in the FORS2 spectrum (Fig. \ref{fors}).  
No studies on Wray 16-30 have been published recently so that its classification as a PN is still debated, though (Reynoso et al.\ 2006 and references therein). Optical spectroscopy is needed to confirm that it is indeed a PN and validate our conclusion.  Unfortunately, because of the observational requirement of having a close-by and reasonably bright reference star to be used for blind offset (Sectn.\ 2.1.1)  we could not place Wray 16-30 within the FORS2  slit. This object has a complex morphology,  resolved by our WFPC2 images into a point-like source and a clumpy structure aligned in the northeast/southwest direction (see Fig.\ 2 of Mignani et al.\ 2009b). However, these are not resolved in our lower-spatial resolution Omegacam images (Sectn.\ 3.2), so that we cannot obtain conclusive information on the  Wray 16-30  spectrum from broad-band photometry alone.

A possible alternative origin of the nebula is that of an emission knot associated with the complex of bright arc-like structures seen 30\arcsec--40\arcsec\ west of it and spatially connected with Wray 16-30.  
Indeed, emission knots, clearly resolved in the Omegacam image (Fig. \ref{vst}a), are seen all along such structures and in their close surroundings. 
These structures coincide with an extended  radio source detected at 13 cm with a thermal continuum spectrum (Reynoso et al.\ 2006) but both their nature and their possible association with Wray 16-30, or with some of the large-scale filaments observed in the field (Fig.\ref{vst}a), are uncertain. Parts of these arc-like structures, however,  fall in the FORS2  slit  (Fig. \ref{fors2}a) so that we can determine their spectrum and compare it with that of the nebula. We extracted spectra from two different regions close to the edges of the structures   (Fig. \ref{fors2}a)  and selected from the target acquisition and thru-slit images. For the spectrum extraction we followed the same approach employed for the nebula (Sectn.\ 2.1.2), i.e.  we used a rectangular window of 1\arcsec\ width and sampled the sky background in a region free of contamination from bright point-like or extended sources. The spectra of the two regions  in the wavelength interval around 6500 \AA\ show the [NII] doublet and the H$\alpha$ line (Fig. \ref{fors2}b). The former is much weaker than in the nebula spectrum though, whereas the latter is much stronger, which suggests different element abundances and/or physical conditions (e.g. gas density, temperature) in the three sampled regions. This can also be reflected by the difference in morphology and surface brightness observed in the Omegacam narrow-band image of these structures (see Sectn.\ 3.2.1). Overall, however, the three spectra are qualitatively similar so that the hypothesis that the nebula is indeed an emission knot from these large structures seems plausible.

Investigating in more detail these and other possible origins of the nebula once evidence suggests  that it is not 
associated with \cxo\ is beyond the goals of our work and we will not discuss this subject any further. Dedicated spectroscopy observations to spatially map the spectra of the arc-like structures west of it, as well as of the  large-scale filaments east/south-east and of the candidate PN Wray 16-30 are needed to determine their properties and verify one hypothesis against the other.

\section{Summary and conclusions}

We have carried out a multi-wavelength follow-up of the CCO \cxo\ in the Vela Jr. SNR and of its tentatively associated emission-line nebula with the  VLT, the VST and \chan. 
Our new VLT near-infrared images, obtained about ten years after those of Mignani et al.\ (2007), do not show evidence of a proper motion of the candidate CCO counterpart, with a  $3 \sigma$ limit of 10 mas yr$^{-1}$. This is much lower than the  proper motion of $\sim$ 50--100 mas yr$^{-1}$ expected  for a SNR age between 2.4 and 5.1 kyr (Allen et al.\ 2015) and the 4\arcmin\ offset between the CCO position and the  estimated centre of the SNR (Aschenbach 1998).  Therefore, 
evidence seem to point against  the identification of \cxo\ with the proposed counterpart,  although it cannot be firmly ruled out yet. 
By comparing the  \cxo\  positions measured in two \chan\  images obtained in 2000 and 2009 we set a $3 \sigma$ upper limit of 300 mas yr$^{-1}$ on its proper motion,  the first constrain directly obtained so far, corresponding to a transverse velocity $V_{\rm T} \la 1500$ km s$^{-1}$ for a SNR distance of 1 kpc (Slane et al.\ 2001; Allen et al.\ 2015).
Through VST H$\alpha$ imaging we confirmed the existence of the nebula around \cxo\ observed in archival data by Pellizzoni et al.\ (2002).  However, VLT spectroscopy revealed that the nebula emission around 6500 \AA\ is resolved in the [NII] doublet at 6548 \AA\ and 6584 \AA, with a marginal contribution from  H$\alpha$ at 6562.8 \AA. Therefore, the  most obvious interpretation of the nebula as a velocity driven bow-shock associated with the CCO is ruled out. Like the other CCOs, \cxo\ remains unidentified at energies other than the X-rays.

\section*{Acknowledgments}

We thank the anonymous referee for his/her constructive comments to our manuscript. RPM  thanks Dr. Maryam Habibi (MPE) for useful discussion on astrometry with the NACO instrument and Dr. Lowell Tacconi-Garman (ESO) for support during the execution of  NACO program 098.D-0346(A).

\label{lastpage}

\end{document}